\documentclass[twocolumn,prl,amsmath,amssymb,superscriptaddress,noeprint]{revtex4-1}

\usepackage[utf8]{inputenc}
\usepackage{hyperref}
\usepackage{amsfonts,amsmath,amssymb}
\usepackage{graphicx}
\usepackage{bm} 
\usepackage{mathrsfs}
\usepackage{subfigure}
\usepackage{textcomp}
\usepackage{hyperref}
\usepackage[flushleft]{threeparttable}
\usepackage[per-mode=symbol]{siunitx}
\usepackage[version=3]{mhchem}

\usepackage{amsmath,amssymb,latexsym}
\usepackage{graphicx}
\usepackage{bm}
\usepackage{mathrsfs}
\usepackage{physics}
\usepackage{siunitx}
\usepackage{mhchem}
\usepackage[normalem]{ulem}
\usepackage{color}
\usepackage{nicefrac}
\usepackage{todonotes}
\usepackage{multirow}
\usepackage{hyperref}
\usepackage[normalem]{ulem}
\usepackage{multirow}

\newcommand{\beq}{\begin{equation}}
\newcommand{\eeq}{\end{equation}}
\newcommand{\beqa}{\begin{eqnarray}}
\newcommand{\eeqa}{\end{eqnarray}}

\graphicspath{{../figures/}}

\DeclareSIUnit\intensity{\watt\per\centi\meter\squared}
\DeclareSIUnit\fieldstrength{\volt\per\centi\meter}
\usepackage{xspace}

\DeclareUnicodeCharacter{2009}{\,}

\newcommand{\cost}{\ensuremath{\langle\cos^2\theta_\text{2D}\rangle}}

\newlength{\figwidth}
\newlength{\figwidthwide}
\let\orgautoref\autoref
\providecommand{\Autoref}{%
   \def\equationautorefname~##1\null{Equation~(##1)\null}%
  \def\figureautorefname{Figure}%
  \def\subfigureautorefname{Figure}%
  \orgautoref}
\renewcommand{\autoref}{%
   \def\equationautorefname~##1\null{Eq.~(##1)\null}%
  \def\figureautorefname{Fig.}%
  \def\subfigureautorefname{Fig.}%
  \orgautoref}

\usepackage{color}
\usepackage[normalem]{ulem}
\definecolor{darkgreen}{rgb}{0.0,0.7,0.0}

\begin{document}

\title{Excited rotational states of molecules in a superfluid}

\author{Igor N. Cherepanov}
\affiliation{Institute of Science and Technology Austria, Am Campus 1, 3400 Klosterneuburg, Austria}

\author{Giacomo Bighin}
\affiliation{Institute of Science and Technology Austria, Am Campus 1, 3400 Klosterneuburg, Austria}

\author{Constant A. Schouder}
\affiliation{Department of Chemistry, Aarhus University, 8000 Aarhus C, Denmark}

\author{Adam S. Chatterley}
\affiliation{Department of Chemistry, Aarhus University, 8000 Aarhus C, Denmark}

\author{Simon H. Albrechtsen}
\affiliation{Department of Chemistry, Aarhus University, 8000 Aarhus C, Denmark}

\author{Alberto Vi\~{n}as Mu\~{n}oz}
\affiliation{Department of Chemistry, Aarhus University, 8000 Aarhus C, Denmark}

\author{Lars Christiansen}
\affiliation{Department of Chemistry, Aarhus University, 8000 Aarhus C, Denmark}

\author{Henrik Stapelfeldt}
\email[Corresponding author: ]{henriks@chem.au.dk}
\affiliation{Department of Chemistry, Aarhus University, 8000 Aarhus C, Denmark}

\author{Mikhail Lemeshko}
\email[Corresponding author: ]{mikhail.lemeshko@ist.ac.at}
\affiliation{Institute of Science and Technology Austria, Am Campus 1, 3400 Klosterneuburg, Austria}

\date{\today}

\begin{abstract}

We combine experimental and theoretical approaches to explore excited rotational states of molecules embedded in helium nanodroplets using \ce{CS2} and \ce{I2} as examples. Laser-induced nonadiabatic molecular alignment is employed to measure spectral lines for rotational states extending beyond those initially populated at the 0.37 K droplet temperature. We construct a simple quantum mechanical model, based on a linear rotor coupled to a single-mode bosonic bath, to determine the rotational energy structure in its entirety. The calculated and measured spectral lines are in good agreement. We show that the effect of the surrounding superfluid on molecular rotation can be rationalized by a single quantity -- the angular momentum, transferred from the molecule to the droplet.

\end{abstract}

\maketitle

Perhaps the most intriguing example of quantized rotational motion of molecules in a liquid is that of molecules embedded in nanometer-sized droplets of superfluid helium. Infrared absorption spectroscopy of many different molecules has established that the rotational energy structure is similar to that of gas phase molecules. In particular, for linear molecules the rotational energy, $E_J$, can be expressed as~\cite{Nauta2001,GrebenevOCS,nauta_vibrational_2001}:
\begin{equation}
E_J= B^* J(J+1) - D^*J^2(J+1)^2.
\label{eq:B*D*}
\end{equation}
 Here $J$ is the rotational angular momentum of the molecule and $B^\ast$ and $D^\ast$ are, respectively, the rotational and centrifugal distortion constants, that assume ``renormalized''  values compared to their gas-phase counterparts, $B$ and $D$. Theoretical models have been developed to describe renormalization of the spectroscopic constants~\cite{Hartmann1995, Lee1999, Kwon1999, CallegariPRL99, GrebenevOCS, LehmannJCP01, LehmannJCP02, Zillich2004, Zillich:2004cta, LemeshkoDroplets16} and led to the following insights: (i) A molecule co-rotates with a nonsuperfluid He density component resulting in a larger moment of inertia and hence a smaller rotational constant compared to that of the isolated molecule.
(ii) When $J$ increases, the effective He--molecule coupling increases~\cite{LehmannJCP01}, which leads to the $J$--dependent decrease of the rotational constant, $D^*(J+1)$, cf.~\autoref{eq:B*D*}, with $D^\ast$ typically 10$^2$--10$^4$ times larger than $D$~\cite{Choi2006}.

The experimental knowledge of the rotational structure of molecules in He droplets has been obtained almost exclusively from IR spectroscopy. Due to selection rules, IR spectroscopy can only provide information about $E_J$ of those rotational levels that are initially populated~\footnote{Actually one level higher than the highest significantly populated level, due to selection rules $\Delta J = \pm 1$.}.
 At 0.37 K temperature of the droplets, this is the lowest-lying rotational states. Consequently, the question of what happens to $E_J$ when $J$ is increased beyond the values populated at 0.37 K remains unanswered by experiments. Notably, \autoref{eq:B*D*} is not of any use in this regime as it predicts that $E_J$ will start decreasing already when $J$ exceeds a rather small value (e.g. $J \approx 10$ for OCS molecules), and even become negative if $J$ is further increased ($J \gtrsim 14 $ for OCS).
Furthermore, this question has, to our knowledge, also not been answered by theory. Here, we combine theoretical and experimental studies to get insight into rotationally excited states of molecules in He nanodroplets.

Experimentally, we form rotational wave packets~\cite{Stapelfeldt:03,fleischer_molecular_2012} in the molecules by a ps alignment pulse and measure the resulting time-dependent degree of alignment. Fourier transformation of such traces reveals the energy of the rotational states in the wave packets -- a technique established for gas phase molecules~\cite{felker_rotational_1992,riehn_high-resolution_2002,schroter_crasy:_2011,chatterley_laser-induced_2020} and recently demonstrated for He-solvated molecules in the limit of weak alignment pulses~\cite{chatterley_rotational_2020}. We conduct measurements for several alignment pulse intensities to systematically map out the rotational energy structure and thereby circumvent the selection rule limitations of IR spectroscopy.

The experimental setup is almost the same as that used in~\cite{chatterley_rotational_2020}. Briefly, helium nanodroplets, doped with at most one \ce{CS2} or  \ce{I2} molecule are irradiated by two linearly polarized laser beams inside a velocity map imaging spectrometer. The 2.3 ps pulses in the alignment beam set the molecules into rotation. Their time-dependent alignment is measured through Coulomb explosion, induced by the 40 fs pulses in the probe beam, and detection of 2D velocity images of \ce{S+} or \ce{IHe+} ions, respectively. Each probe pulse is delayed, $t$, with respect to an alignment pulse. From the ion images, the degree of the molecular alignment, \cost, is determined, $\theta_\text{2D}$  being the angle between the alignment pulse polarization and the projection of the velocity vector of a \ce{S+} or \ce{IHe+} ion on the detector. For details see \cite{sup}, which includes Refs.~\cite{ToenniesAngChem04,Shepperson17,Sondergaard:2017id}.

\begin{figure}[t]
\centering
\includegraphics[width=1\linewidth]{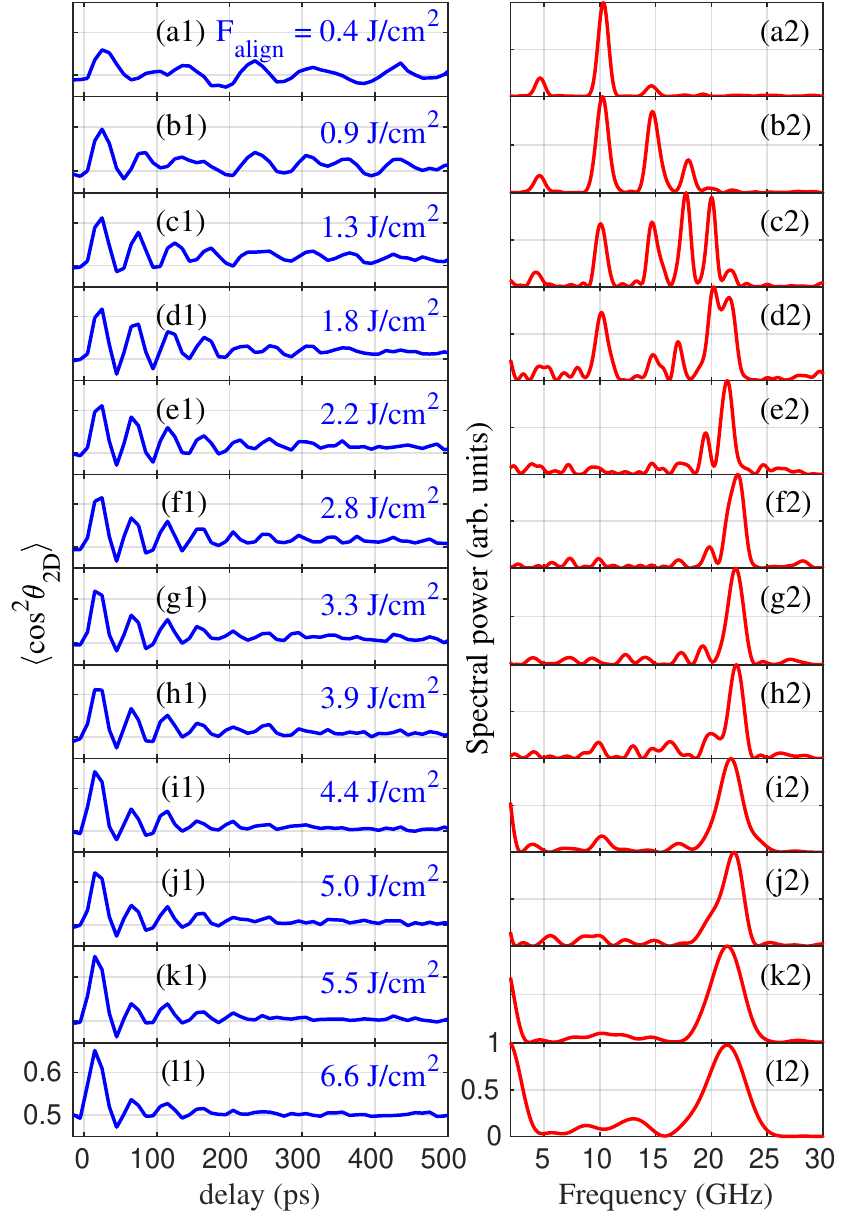}
\caption{Column 1: \cost(t) for \ce{CS2} molecules at 12 different fluences of the alignment pulse given on each panel. Column 2: The power spectra of the corresponding \cost traces. }
\label{fig:CS2}
\end{figure}

Column 1 in \autoref{fig:CS2} shows \cost(t) for \ce{CS2} molecules at different fluences of the alignment pulse and column 2 the corresponding power spectra obtained by Fourier transformation of the \cost~traces~\footnote{The \cost~traces were actually recorded out to 1200 ps. However, in order to exclude noise the power spectra in Fig. 1 were computed using a Hamming window over only the signal bearing part of the \cost~traces. This window ranged from the full 1200 ps trace for the lowest fluences, to the first 400 ps for the highest fluence.}. At the lowest fluences, clearly separated peaks in the spectra are observed. When the fluence, $F$, is increased, the spectrum shifts to higher frequencies, consistent with the expectation that a stronger pulse adds more rotational angular momentum to the molecule, and the distance between the peaks in the high-end part of the spectrum diminishes and start overlapping. For $F\geq$ \SI{2.8}{J/cm^2}, the spectra are dominated by a single broad peak centered at $\sim\SI{22}{GHz}$.

\begin{figure}[t]
\centering
\includegraphics[width=1.0\linewidth]{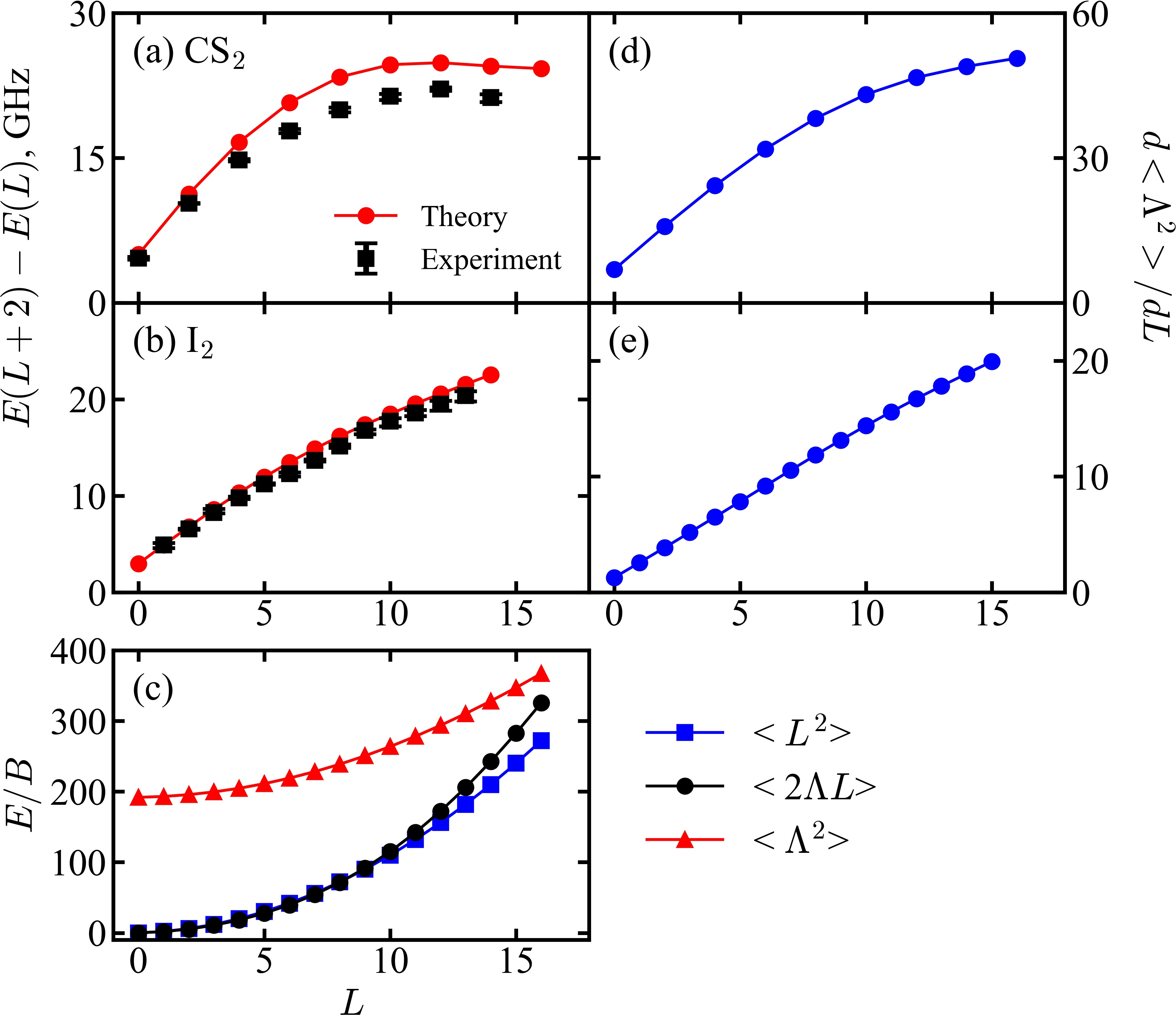}
\caption{(a)-(b): Frequency differences, $(E_{L+2} - E_{L})/h$, as a function of $L$. The black squares are the experimental results obtained from the central positions of the peaks in the spectra, cf.~Fig.~\ref{fig:CS2}, and the red circles are the results from the theoretical model. (c): The three contributions to the rotational energy of I$_2$ molecules in He droplets. (d),(e) First derivative of the He solvent angular momentum squared, $d\langle \mathbf{\Lambda}^2 \rangle/dL$, for CS$_2$ and I$_2$, respectively.}
\label{fig:comparison}
\end{figure}

The series of spectra recorded at incrementally increasing values of $F$ allows us to identify a total of eight peaks. They reflect the frequencies of the $L\leftrightarrow L+2$ coherences of the rotational wave packets~\cite{chatterley_rotational_2020}, where $L$ represents the total angular momentum. In \autoref{fig:comparison}(a), the peak positions, given by $(E_{L+2} - E_{L})/h$~\cite{chatterley_rotational_2020}, are plotted as a function of $L  = 0,2,\dots, 14$. Similarly we identified  $13$ peaks in the spectra of  \ce{I2} \cite{sup}. Their central positions are also plotted as a function of $L = 1,2,\dots,13$, as shown in \autoref{fig:comparison}(b).

For our theoretical model we start from general considerations that are neither dependent on the molecular species nor on the details of the molecule--solvent interactions. Considering only addition of angular momenta, the rotational part of the Hamiltonian for a linear-rotor molecule in a solvent can be written as:
\begin{equation}
\label{eq:RotH}
\hat{H}_\text{rot}=B \boldsymbol{\mathrm{\hat{J}}}^2 = B(\boldsymbol{\mathrm{\hat{L}}}-\boldsymbol{\mathrm{\hat{\Lambda}}})^2
\end{equation}
Here $B$ is the gas-phase rotational constant of the molecule, $\boldsymbol{\mathrm{\hat{J}}}$ is the rotational angular momentum of the molecule, $\boldsymbol{\mathrm{\hat{\Lambda}}}$ is  the angular momentum carried by the surrounding solvent and  $\boldsymbol{\mathrm{\hat{L}}}$ is the total angular momentum, which is the only conserved quantity in the presence of molecule--solvent interactions.

The Hamiltonian \eqref{eq:RotH} can be mapped onto an effective symmetric-top Hamiltonian similar to that of linear open-shell molecules, such as OH or NO~\cite{LevebvreBrionField2}, with the solvent angular momentum $\boldsymbol{\mathrm{\hat{\Lambda}}}$ playing the role of the electron angular momentum. The corresponding states can be expressed through the symmetric top states, $\ket{L n M}$, where $n$ and $M$ label the projection of the total angular momentum $\boldsymbol{\mathrm{\hat{L}}}$ on the molecular and laboratory axes, respectively~\footnote{Note that for a linear molecule, $n$ corresponds to the projection of $\boldsymbol{\mathrm{\hat{\Lambda}}}$ on the molecular axis, since the projection of $ \boldsymbol{\mathrm{\hat{J}}}$ vanishes.}.

From the right-hand side of \autoref{eq:RotH} one can see that the energies corresponding to a given total angular momentum, $L = 0,1,2,\dots$, depend on the value of the solvent angular momentum $\boldsymbol{\mathrm{\Lambda}}$.  In other words, the rotational constant  of a molecule in a solvent is strictly speaking not renormalized. Rather, the change of the rotational energies compared to the gas phase case can be explained by the finite value of $\boldsymbol{\mathrm{\Lambda}}$, see below.

Let us assume that an excitation of the solvent has angular momentum $\lambda$ and energy $\omega$. While in the case of superfluid helium it might be tempting to label these excitations as ``rotons'', we intentionally keep the treatment as general as possible and focus solely on energy and angular momentum conservation. If we initially  include only single excitations of the solvent and consider only the diagonal terms in \autoref{eq:RotH} in the $\ket{L n M}$ basis, the eigenvalues of $\hat{H}_\text{rot}$ are given by:
\begin{equation}
\label{eq:ELn}
E_{L,n}= BL(L+1) -2Bn^2 +B\lambda(\lambda+1) + \omega,
\end{equation}
where we introduced an additional energy shift by the excitation energy $\omega$. $E_{L,n}$ is independent of $M$ in the absence of external fields.  \Autoref{eq:ELn} corresponds to an oblate symmetric top  shifted by $B\lambda(\lambda+1) + \omega$ from the zero energy, whose eigenstates are given by $\ket{L n M}$.  When the off-diagonal terms of $\hat{H}_\text{rot}$ are taken into account, $\ket{L n M}$ states with different $n$'s are mixed. This alters the eigenenergies compared to  \autoref{eq:ELn} and renders $n$ an approximate quantum number. In what follows, we retain $n$ to label the eigenenergies $E_{L,n}$.

\begin{figure}[t]
\centering
\includegraphics[width=1\linewidth]{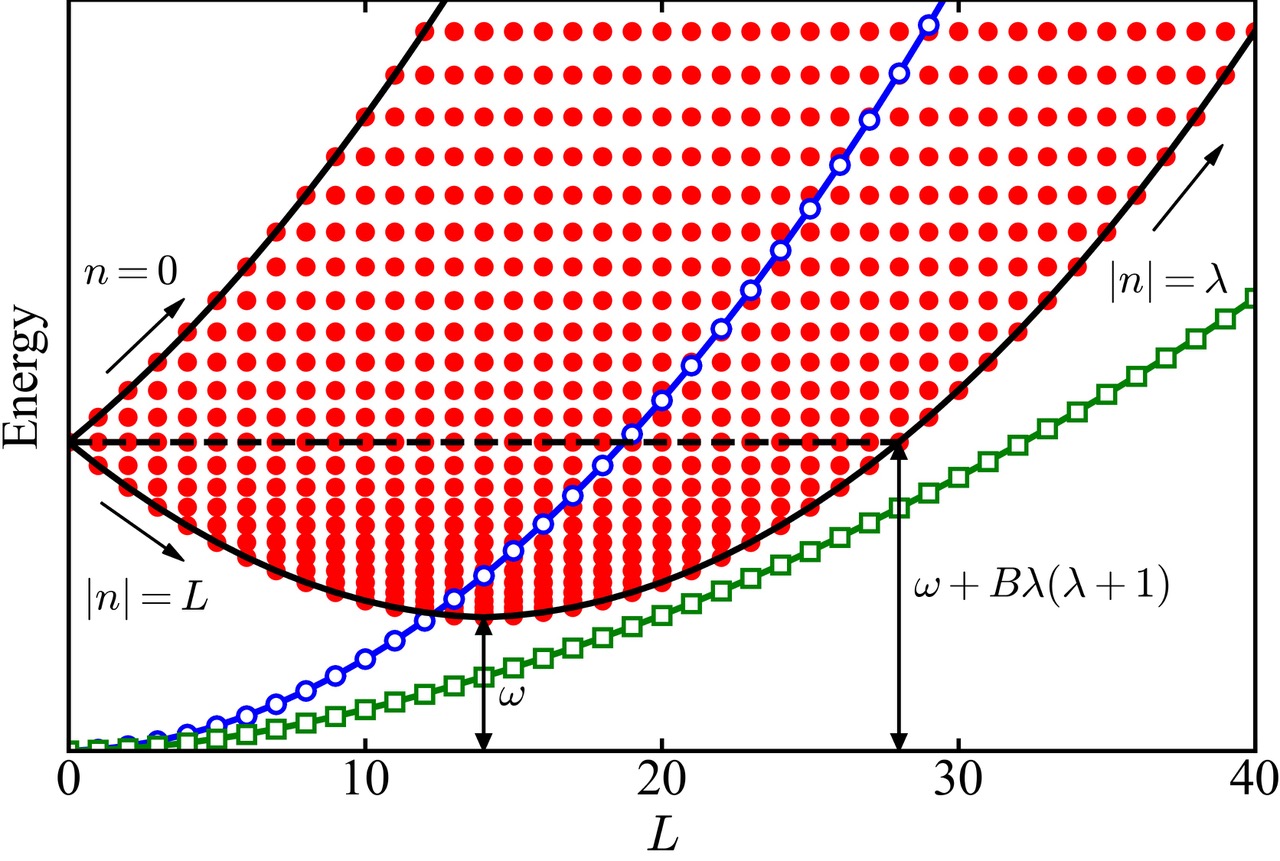}
\caption{The gas-phase molecular rotational states (blue circles) are perturbed by a band of excited states (red filled circles, only single solvent excitations are shown), resulting in the rotational states of  molecule in the presence of the solvent (green squares, a calculation for I$_2$ molecule including multiple solvent excitations). }
\label{fig:band}
\end{figure}

The red dots in \autoref{fig:band} show $E_{L,n}$ with off-diagonal terms of $\hat{H}_\text{rot}$ included. For each $L$ (conserved quantity), the $E_{L,n}$'s form a band of 2$n$+1 excited states, originating at $B\lambda(\lambda+1) + \omega$. Due to molecule--solvent interactions, whose exact form is for now irrelevant, these states couple to the free molecular states, $E^{(0)}_{L}= BL(L+1)$ (blue circles). This perturbation results in the rotational states of the He-solvated molecules, shown by green squares. Note that these states were determined by taking into account both single, double and triple solvent excitations. \Autoref{fig:band} shows, however, only states from single excitations to avoid an unreadable figure. The data presented in \autoref{fig:band} were calculated for the I$_2$ molecule but the qualitative features of the energy of the states are general.

The shape of the band of excited states is determined by the range of the possible projections $-L \leq n \leq L$ (for a linear molecule and single excitation, $|n| \leq \lambda$ must also be fulfilled). In the language of perturbation theory, the closer the band edge is to the blue line, the stronger is the free molecular state perturbed by the solvent excitations and the more the green squares deviate from the gas-phase behavior. The perturbation is the strongest after the free molecular states cross the excitation threshold at energy $\sim \omega$, which happens for $L\gtrsim 12$, cf. \autoref{fig:comparison}(c).

Let us consider a  linear rotor coupled to a bosonic bath with a single mode carrying energy $\omega$ and angular momentum $\lambda$, as described by the following Hamiltonian  in the molecular frame \cite{SchmidtLem16}:
\begin{equation}
\label{eq:Hamiltonian}
\hat{H}=B(\boldsymbol{\mathrm{\hat{L}}}-\boldsymbol{\mathrm{\hat{\Lambda}}})^2 +  \omega \sum_{\mu} \hat{b}^{\dagger}_{\lambda \mu}\hat{b}_{\lambda \mu}+ u \big(\hat{b}^{\dagger}_{\lambda 0} +\hat{b}_{\lambda 0}\big) \;
\end{equation}
where $u$ gives the molecule--solvent interaction strength, $\hat{b}^{\dagger}_{\lambda \mu}$($\hat{b}_{\lambda \mu}$) create (annihilate) a solvent excitation with angular momentum $\lambda$ and projection onto the molecular axis $\mu$, and $\boldsymbol{\mathrm{\hat{\Lambda}}}=\sum_{\mu \nu} \hat{b}^{\dagger}_{\lambda \mu} \boldsymbol{\mathrm{\sigma}}^{\lambda}_{\mu \nu}\hat{b}_{\lambda \nu}$ is the angular momentum of the solvent.

We diagonalize the Hamiltonian in the basis containing multiple excitations of the single bosonic mode and all possible projections $n$:
  \begin{equation}
\psi_{L [n_1 n_2...n_m], M}^{(m)} =  \ket{LNM}_\text{mol} \left( b^{\dagger}_{\lambda n_1} b^{\dagger}_{\lambda n_2} ...  b^{\dagger}_{\lambda n_m}\ket{0}_\text{bos} \right)
\label{eq:psi}
\end{equation}
where $N = \sum  n_i$ (with $\vert N \vert \leq L$) is the total angular momentum projection of the bosons. We take into account the states with $m \leq 3$, however, the arguments  used for the single-excitation case of ~\autoref{fig:band} above, still hold in this case. Most important, the asymptotic behavior of the excited state band at $L \to \infty$ is identical to that shown in~\autoref{fig:band}.

We used the model
to calculate the rotational energies of \ce{CS2} and \ce{I2} molecules in He droplets (the values of the parameters  are listed in~\cite{sup}). Green squares in \autoref{fig:band} represent the results for \ce{I2}. For comparison with experiment, the frequency differences, $(E_{L+2} - E_{L})/h$, were determined and plotted in Figs.~\ref{fig:comparison}(a)-(b) by red circles. For both \ce{I2} and \ce{CS2} there is good agreement between the experimental and theoretical results. Notably, the model predicts that the difference spectrum exhibits a maximum frequency around  $\SI{20}-\SI{25}{GHz}$ very similar to that observed in experiment. Such a maximum never occurs for a free rotor where the position of a spectral line, $(E_{L+2} - E_{L})/h = 4L +6$, increases linearly with $L$~\footnote{Only at extreme values of $L$, reachable by an optical centrifuge will centrifugal distortion cause a deviation from the linear behaviour \cite{milner_probing_2017}.}.

In what follows, we show how the rotational energy structure of molecules in He droplets can be rationalized in terms of a single quantity -- the angular momentum of the solvent~$\boldsymbol{\mathrm{\Lambda}}$. \Autoref{fig:comparison}(c) displays the $L$-dependence of $B\langle \mathbf{L}^2 \rangle$, $B\langle \mathbf{\Lambda}^2 \rangle$ and $B\langle 2 \mathbf{\Lambda \cdot L} \rangle$, the three contributions to the molecular rotational energy cf. \autoref{eq:RotH}, calculated for I$_2$ molecules (the curves for CS$_2$ show a similar behavior~\cite{sup}).

\begin{figure}[t]
\centering
\includegraphics[width=0.9\linewidth]{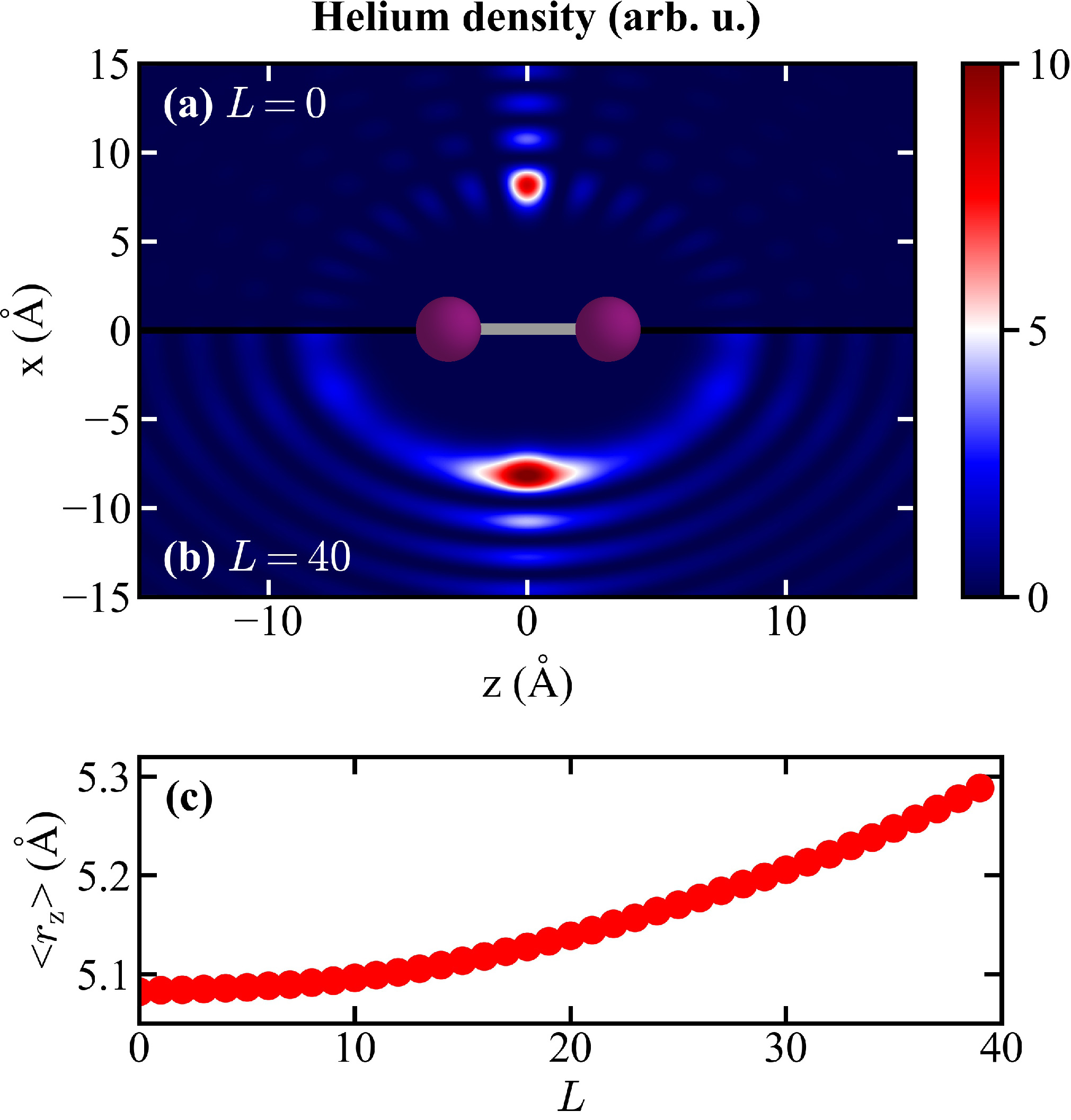}
\caption{Anisotropic component of the helium density in the molecular frame for (a) $L=0$ and (b) $L=40$. (c) The average distance of He atoms from the $z=0$ axis grows with the angular momentum $L$. All for \ce{I2} molecule.}
\label{fig:densities}
\end{figure}

For $L=0$, a non-rotating molecule is ``dressed'' by the excitations of the  solvent due to molecule--He interactions. The corresponding many-particle state is given by a superposition of the basis states \autoref{eq:psi} with zero and non-zero angular momentum, which results in a nonzero expectation value $\langle \mathbf{\Lambda}^2 \rangle$, see \autoref{fig:comparison}(c).  Note that in order to have $\langle \mathbf{\Lambda}^2 \rangle \neq 0$, the solvent atoms do not need to be physically rotating. Barely placing an anisotropic molecule into the solvent deforms its density in a non-spherically symmetric fashion and thereby provides it with a nonzero angular momentum, cf. \autoref{fig:densities}.

\Autoref{fig:comparison}(c) shows that $\langle \mathbf{L}^2 \rangle \approx \langle 2 \mathbf{\Lambda \cdot L} \rangle$ up to $L \sim 10$, which implies that $\langle \mathbf{J}^2 \rangle = \langle (\mathbf{L} - \mathbf{\Lambda})^2 \rangle \approx \langle \mathbf{\Lambda}^2 \rangle$, {\it i.e.} the angular momenta of the solvent and of the molecule are equal to each other. Classically this can be understood as follows. At small values of $L$, the He atoms are almost rigidly attached to the molecule and co-rotate with the molecule, which results in the equal magnitude of the molecular and solvent angular momenta. This is analogous to the ``non-superfluid solvation shell'', previously discussed in the literature~\cite{ToenniesAngChem04}. In order for the molecular rotation to be able to perturb the ``solvation shell'', the rotational kinetic energy must be comparable to the energy of the solvent excitations, i.e. $B L(L+1) \sim \omega$, which for I$_2$ molecules corresponds to $L \sim 12$. Around this point we observe the deviations of $\langle 2 \mathbf{\Lambda \cdot L} \rangle$ from $\langle \mathbf{L}^2 \rangle$  in \autoref{fig:comparison}(c).

As we show in Figs.~\ref{fig:comparison}(d)-(e) the derivative of $d\langle \mathbf{\Lambda}^2 \rangle/dL$ alone is able to describe the experimental data on the rotational energy splittings of the CS$_2$ and I$_2$ in superfluid helium, Figs.~\ref{fig:comparison}(a,b), quite accurately. From this plot one can clearly see the origin of the ``renormalized'' spectroscopic constants of \autoref{eq:B*D*}. The linear behavior of $d\langle \mathbf{\Lambda}^2 \rangle/dL$ corresponds to an ``effective $B^\ast$''. The deviations of $d\langle \mathbf{\Lambda}^2 \rangle/dL$ from the linear behavior, together with the contributions from the $\langle 2\mathbf{\Lambda \cdot L}\rangle$ at higher $L$ lead to higher-order terms, such as $D^\ast L^2 (L+1)^2$. Furthermore, while the magnitude of $\langle \mathbf{\Lambda} \rangle$ grows with $L$, it saturates and assumes a constant value  for large $L$~\cite{sup}. As a result,  $\Lambda$ plays a decreasing role for $L\to \infty$, meaning that the rotational spectrum eventually approaches that of the gas phase, which can classically be interpreted as detachment of the molecule from the surrounding superfluid~\cite{Shepperson:2017gb}.

Moreover, the model yields analytic expressions for $B^\ast$ and $D^\ast$ in agreement with experiments~\cite{ToenniesAngChem04}. In the basis of single solvent excitations, the constants can be approximated as \cite{sup}

\begin{equation}
\frac{B^*}{B} \approx 1- \frac{\tilde{u}^2 }{\left( 1 + \tilde{\omega} \right)^3}; \hspace{0.5cm} \frac{D^*}{B} \approx \frac{\tilde{u}^2}{ \lambda(\lambda+1) \left( 1+ \tilde{\omega} \right)^5}
\label{eq: BDapprox}
\end{equation}
with $\tilde{u} = u/[B \lambda(\lambda+1)]$ and $\tilde{\omega}=\omega/[B \lambda(\lambda+1)]$. In the limit of light molecules $(\tilde{u},\tilde{\omega}) \to 0$, $B^\ast$ approaches $B$. Furthermore, \autoref{eq: BDapprox} provides a useful relation, $D^\ast/B \approx \xi \left(1 - B^\ast/B \right)^{5/3}$, with $\xi = \tilde{u}^{-4/3} /[\lambda (\lambda+1) ]$. This dependence is similar to the power law, $D^* =0.031 \cross B^{*1.818}$, found in Ref.~\cite{Choi2006} by fitting the experimental data, however, provides a correct limit of $D^\ast \to 0$ for $B^\ast \to B$.

To illustrate what the change of angular momentum of \autoref{fig:comparison}(c) corresponds to in real space, we used the model to evaluate the helium density distribution in the molecular frame, as shown in Figs.~\ref{fig:densities}(a,b) for $L=0$ and $L=40$, respectively. Note that since our model accounts for He droplet excitations by a single mode only, the model is not expected to provide quantitatively accurate density estimates.  We observe that with the growth of angular momentum $L$, the solvent atoms slightly move from the direction perpendicular to the molecular axis (``T-shape'' configuration) towards the linear configuration, which is clearly visible in the average distance of the He atoms from the $z$-axis, \autoref{fig:densities}(c).  This reflects the fact that the $B^\ast$ and $D^\ast$ constants can be interpreted in terms of formation of the molecule-He-shell complex and the displacement of the He atoms due to rotation of the complex.

In conclusion, through ps time-resolved alignment experiments, we measured the rotational energy levels of \ce{CS2} and \ce{I2} in He droplets up to $L$ = 16 and 15, respectively. The results agree well with the outcome of a quantum model, which also shows that the rotational energy structure can be understood in terms of the angular momentrum transferred from the molecules to the He solvent. Interesting future challenges for both experiments and theory include an understanding of the lifetimes of the highest excited states observed.

\begin{acknowledgments}
I.C.~acknowledges the support by the European Union's Horizon 2020 research and innovation programme under the Marie Sk\l{}odowska-Curie Grant Agreement No.~665385. G.B.~acknowledges support from the Austrian Science Fund (FWF), under project No.~M2461-N27. M.L.~acknowledges support by the Austrian Science Fund (FWF), under project No.~P29902-N27, and by the European Research Council (ERC) Starting Grant No.~801770 (ANGULON). H.S~acknowledges support from the European Research Council-AdG (Project No. 320459, DropletControl) and from The Villum Foundation through a Villum Investigator grant no. 25886.
\end{acknowledgments}

\clearpage
\pagebreak
\widetext

\setcounter{equation}{0}
\setcounter{figure}{0}
\setcounter{table}{0}
\setcounter{page}{1}
\makeatletter
\renewcommand{\theequation}{S\arabic{equation}}
\renewcommand{\thefigure}{S\arabic{figure}}
\renewcommand{\thetable}{S\arabic{table}}

\begin{center}
\textbf{\large Supplemental Material:\\[5pt] Excited rotational states of molecules in a superfluid}
\end{center}

\section{Experimental setup and methods}

Details of the experimental setup were described in~\cite{shepperson_strongly_2017} and, thus, only a brief description of the relevant details are given here. Helium droplets are produced using a continuous helium droplet source with stagnation conditions of \SI{25}{bar} and \SI{14}{K} for \ce{I2} and \ce{CS2}, giving $\sim$\SI{10}{nm} diameter helium droplets~\cite{toennies_superfluid_2004}. The beam of He droplets exit the source chamber through a skimmer with a \SI{1}{mm} diameter opening and enters a pickup cell containing a vapor of either \ce{CS2} or \ce{I2} molecules. The partial pressure of the molecular vapor was kept sufficiently low to ensure the pickup of at most one molecule per droplet. Hereafter, the doped droplets pass through a liquid nitrogen trap that captures the majority of the effusive molecules that are not picked up by the droplets. In order to further reduce the contribution from effusive molecules the doped droplets pass through a second skimmer with a \SI{2}{mm} diameter opening followed by a second liquid nitrogen trap. Finally, the doped droplets enter a velocity map imaging (VMI) spectrometer placed in the middle of the target chamber. Here, the droplet beam is crossed perpendicularly by two collinear, focused, pulsed laser beams, both with a central wavelength of \SI{800}{nm}

The pulses in the first beam are used to induce alignment. These pulses have a duration of \SI{2.3}{ps} and a Gaussian spotsize, $\omega_\text{0}$ = \SI{30}{\micro\metre}. The pulses, termed probe pulses, in the second beam, sent at time $t$ with respect to the center of the alignment pulses, are used to measure the spatial orientation of the molecules. This occurs by Coulomb explosion of the molecules and recording of the direction of the fragment ions recoiling along the internuclear axis of their parent molecule, \ce{IHe^+} for \ce{I_2} and \ce{S^+} for \ce{CS_2}. These pulses have a duration of \SI{40}{fs}, spotsize, $\omega_\text{0}$ = \SI{22}{\micro\metre}, and a peak intensity $\SI{2e14}{W/cm^2}$.

The VMI spectrometer projects the ions produced by the probe pulse onto a 2-dimensional detector. The angle between the position of an ion hit and the polarization direction of the alignment beam, contained in the detector plane, is denoted $\theta_\text{2D}$. The degree of alignment is characterized by $\langle \cos^2\theta_\text{2D} \rangle$, a standard measure used in many previous works~\cite{sondergaard_nonadiabatic_2017}. The 2-dimensional ion images are recorded at a large number of delays between the alignment and the probe pulse.  Hereby the time-dependent $\langle \cos^2  \theta_\text{2D} \rangle$ curves, displayed in Fig. 1 in the main article, are obtained.

\section{Alignment traces and spectra for \ce{I2} molecules }

\begin{figure}[t]
\centering
\includegraphics[width=0.6\linewidth]{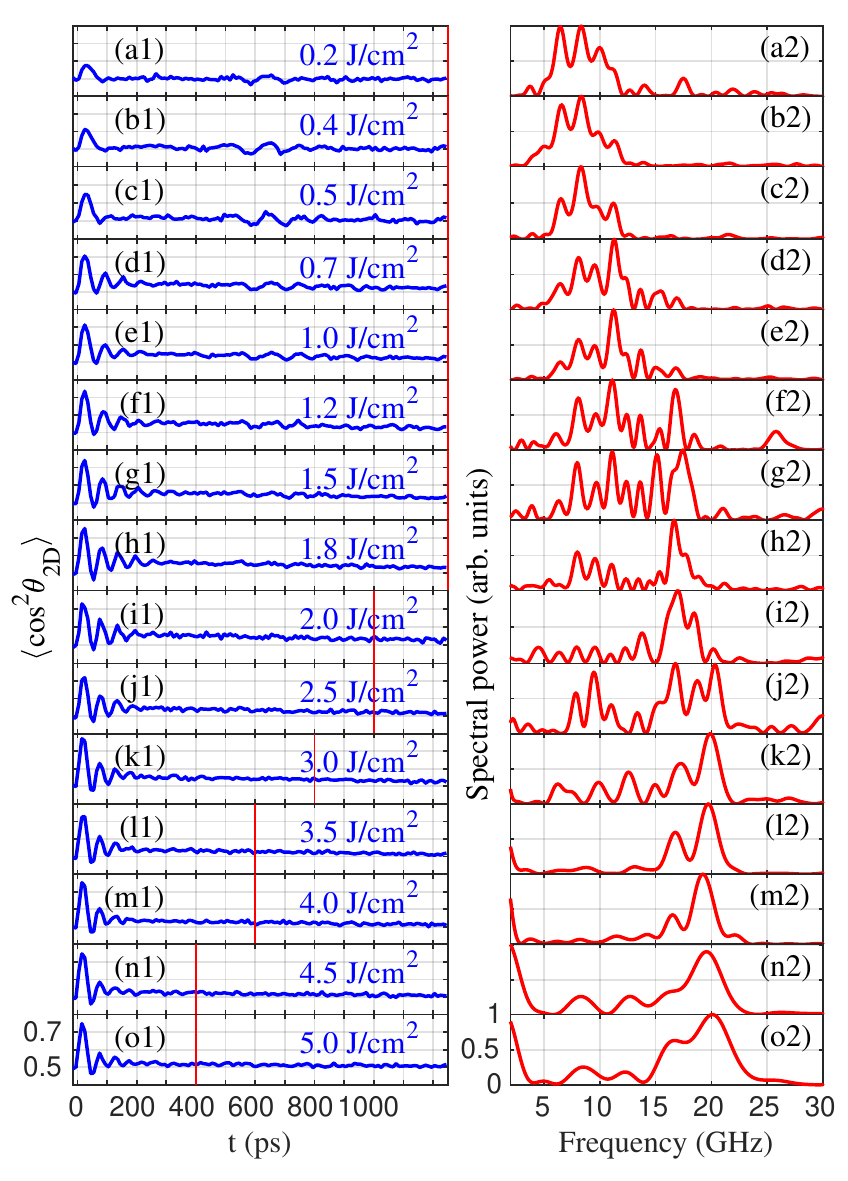}
\caption{Column 1: \cost(t) for \ce{I2} molecules at 15 different fluences of the alignment pulse given on each panel. The red vertical lines indicate the size of the window used to obtain power spectra. Column 2: The power spectra of the corresponding \cost traces. }
\label{fig:I2}
\end{figure}

Column 1 in \autoref{fig:I2} shows \cost(t) for \ce{I2} molecules at different fluences of the alignment pulse and column 2 the corresponding power spectra obtained by Fourier transformation of the \cost traces. As for the data on \ce{CS2}, the \cost traces were recorded out to 1200 ps. To exclude noise the power spectra in \autoref{fig:I2} were computed using a Hamming window over only the signal bearing part of the \cost traces. This window ranged from the full 1200 ps trace for the lowest fluences, to the first 400 ps for the highest fluence.

The observations are qualitatively similar to those made for \ce{CS2}, i.e. clearly separated peaks in the spectra at the lowest fluences, a shift to higher frequencies and overlap between the peaks when $F$ is increased, and eventually merging into a single broad peak, centered at $\sim\SI{18}{GHz}$, at the highest fluences. The series of spectra recorded at incrementally increasing values of $F$ allows us to identify a total of 13 peaks. Their central positions are plotted as a function of $L$, $L = 1,\dots,13$, in Fig. 2(b) in the main text.

\section{Rotational energy in the angulon Hamiltonian}

We provide the explicit expression for the rotational Hamiltonian in Eq.~(2) of the main text:
\begin{equation}
\label{eq:Hamiltonian}
\hat{H}=B(\boldsymbol{\mathrm{\hat{L}}}-\boldsymbol{\mathrm{\hat{\Lambda}}})^2 = B(\boldsymbol{\mathrm{\hat{L}}}^2 - 2( \mathrm{\hat{L}_0} \mathrm{\hat{\Lambda}_0} - \mathrm{\hat{L}_-} \mathrm{\hat{\Lambda}_+} - \mathrm{\hat{L}_+} \mathrm{\hat{\Lambda}_-}) + \boldsymbol{\mathrm{\hat{\Lambda}}}^2)
\end{equation}

The angular-momentum operator for the bosonic helium bath $\boldsymbol{\mathrm{\hat{\Lambda}}} = \{\mathrm{\hat{\Lambda}_-}, \mathrm{\hat{\Lambda}_0},\mathrm{\hat{\Lambda}_+} \}$ is given by 

\begin{equation}
\boldsymbol{\mathrm{\hat{\Lambda}}}=\sum_{\mu \nu} \hat{b}^{\dagger}_{\lambda \mu} \boldsymbol{\mathrm{\sigma}}^{\lambda}_{\mu \nu}\hat{b}_{\lambda \nu}
\end{equation}
where $\boldsymbol{\mathrm{\sigma}}^{\lambda}=\{\sigma^{\lambda}_{-1},\sigma^{\lambda}_{0},\sigma^{\lambda}_{+1}\}$ denotes the vector of the angular momentum matrices fulfilling the $SO(3)$ algebra in the representation of angular momentum $\lambda$ \cite{SchmidtLem16_sup}. In the basis of single bosonic excitations $\psi^{(1)}_{LMn} = \vert LMn \rangle_{\mathrm{mol}} b^{\dagger}_{\lambda n} \vert 0 \rangle_{\mathrm{bos}}$, the corresponding matrix elements read:

\begin{equation}
\langle \psi^{(1)}_{LMn} \vert \boldsymbol{\mathrm{\hat{L}}}^2 \vert \psi^{(1)}_{LMn'}  \rangle = L(L+1) \delta_{n,n'}
\end{equation}

\begin{equation}
\langle \psi^{(1)}_{LMn} \vert -\mathrm{\hat{L}_-} \mathrm{\hat{\Lambda}_+}  \vert \psi^{(1)}_{LMn'}  \rangle = \frac{1}{2} \sqrt{\lambda(\lambda+1)-\nu(\nu+1)}\sqrt{L(L+1)-\nu(\nu+1)} \delta_{n,n'+1}
\end{equation}

\begin{equation}
\langle \psi^{(1)}_{LMn} \vert -\mathrm{\hat{L}_+} \mathrm{\hat{\Lambda}_-}  \vert \psi^{(1)}_{LMn'}  \rangle = \frac{1}{2} \sqrt{\lambda(\lambda+1)-\nu(\nu-1)}\sqrt{L(L+1)-\nu(\nu-1)} \delta_{n,n'-1}
\end{equation}

\begin{equation}
\langle \psi^{(1)}_{LMn} \vert \mathrm{\hat{L}_0} \mathrm{\hat{\Lambda}_0}  \vert \psi^{(1)}_{LMn'}  \rangle = n^2 \delta_{n,n'}
\end{equation}

\begin{equation}
\langle \psi^{(1)}_{LMn} \vert \boldsymbol{\mathrm{\hat{\Lambda}}}^2 \vert \psi^{(1)}_{LMn'}  \rangle = \lambda(\lambda+1) \delta_{n,n'}
\end{equation}

\section{Derivation of estimates for $B^*$ and $D^*$}

To obtain simple estimates for $B^*$ and $D^*$, we introduce a variational ansatz accounting for single bosonic excitations:

\begin{equation}
\vert \Psi_{LM} \rangle = g \vert  \psi^{(0)} \rangle +\beta_{\lambda n } \vert  \psi^{(1)}_{LMn} \rangle  = g \vert LM0 \rangle_{\mathrm{mol}} \vert  0 \rangle_{\mathrm{bos}} +\beta_{\lambda n } \vert LMn \rangle_{\mathrm{mol}} b^{\dagger}_{\lambda n} \vert 0 \rangle_{\mathrm{bos}}
\end{equation}
where $g$ and $\beta_{\lambda n}$ are variational coefficients.

\par

Minimization of the energy functional $F = \langle \Psi_{LM} \vert E - \hat{H} \vert \Psi_{LM}  \rangle$ with respect to variational coefficients yields:

\begin{equation}
    \begin{cases}
     	g\bigg(-E+BL(L+1)\bigg) + u \beta_{n}  \delta _{n,0} ~=0   \\
\beta_{n} \bigg(-E+ BL(L+1)+B\lambda(\lambda+1) + \omega \bigg) -2B \sum_{n'}  \beta_{n'}  \boldsymbol{\eta}^L_{n,n'} \boldsymbol{\sigma}^{\lambda}_{n,n'} -u g \delta _{n,0} ~=0
    \end{cases}\
    \label{system}
\end{equation}
where $\boldsymbol{\eta}^L_{n,n'} \boldsymbol{\sigma}^{\lambda}_{n,n'} = \langle \psi^{(1)}_{LMn} \vert \mathrm{\hat{L}_-} \mathrm{\hat{\Lambda}_+}  \vert \psi^{(1)}_{LMn'}  \rangle + \langle \psi^{(1)}_{LMn} \vert \mathrm{\hat{L}_+} \mathrm{\hat{\Lambda}_-}  \vert \psi^{(1)}_{LMn'}  \rangle + \langle \psi^{(1)}_{LMn} \vert \mathrm{\hat{L}_0} \mathrm{\hat{\Lambda}_0}  \vert \psi^{(1)}_{LMn'}  \rangle $.

\par

Since $B^*$ and $D^*$ are used to fit energies of low-lying rotational states, we will restrict our consideration to two lowest projections $n$ in \autoref{system}, this is exact for $L=1$. In the framework of such an approximation, we obtain the following solution of the system:
 
\begin{equation}
E  = BL(L+1) - \frac{u^2 E_1}{E_0 E_1 - B^2 L(L+1) \lambda (\lambda+1)}
\end{equation}
where $E_0 = -E+BL(L+1)+B \lambda(\lambda+1)+\omega$ and $E_1 = -E+BL(L+1)+B \lambda(\lambda+1)+\omega - 2B$. Taking into account that $-E+BL(L+1) - 2B << \omega+ B \lambda(\lambda+1)$ and expanding the energy in powers of $L(L+1)$, we finally get:

\begin{equation}
E  = BL(L+1) - u^2 \bigg(\frac{1}{\omega+ B \lambda(\lambda+1)} + \frac{B^2 L(L+1)\lambda(\lambda+1)}{(\omega+ B \lambda(\lambda+1))^3} + \frac{B^4 L^2(L+1)^2\lambda^2(\lambda+1)^2}{(\omega+ B \lambda(\lambda+1))^5 } + O(L^3(L+1)^3)\bigg)
\end{equation}

\section{Breakdown of the rotational energy for \ce{CS_2}}

\begin{figure}[h!]
\hspace*{2.3cm}
\includegraphics[width=0.62\linewidth]{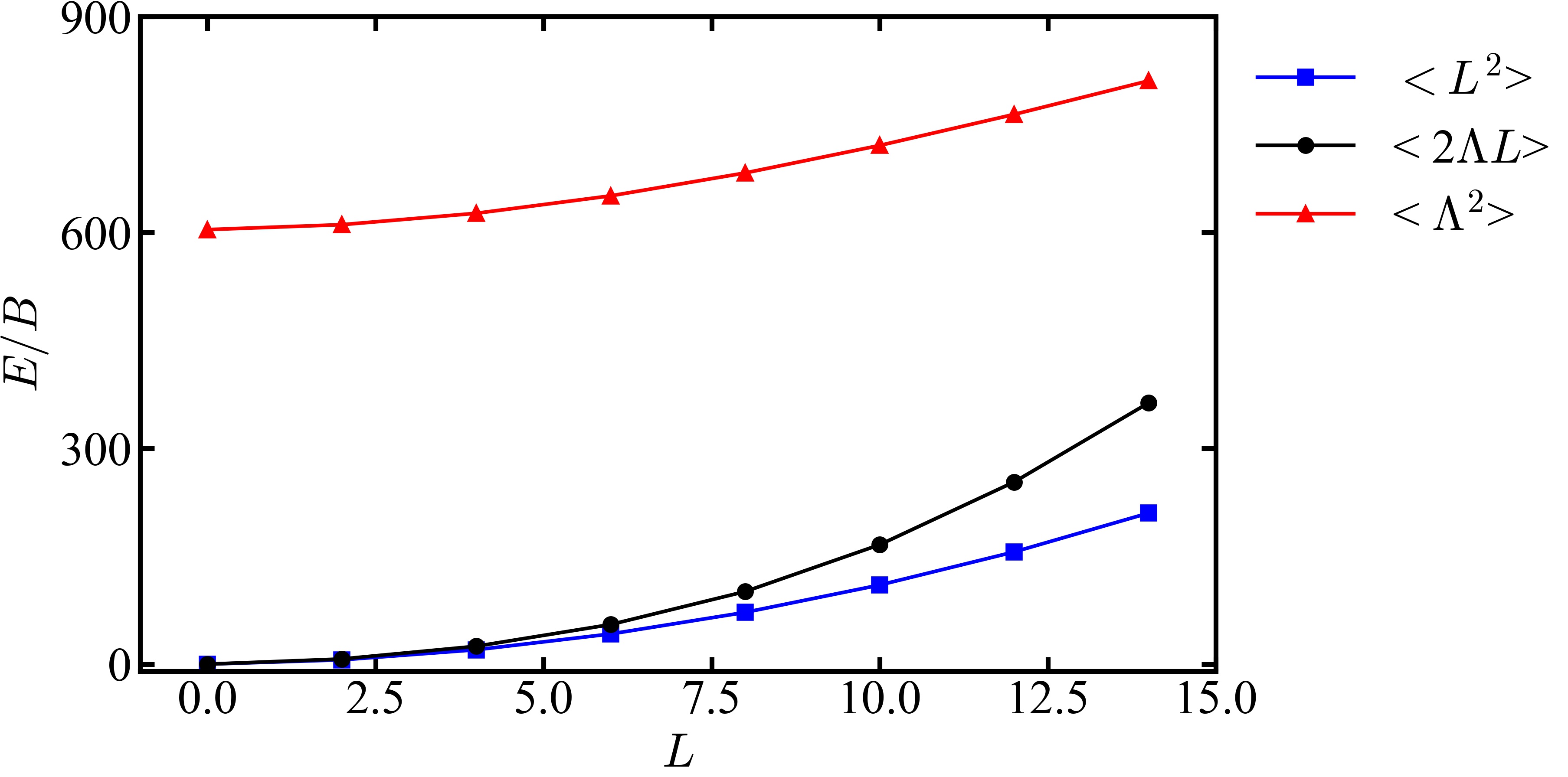}
\caption{ Different contributions to the rotational energy of \ce{CS_2} molecules in He droplets. }
\label{s2}
\end{figure}

\Autoref{s2} is analogous to Fig.2(c) of the main text. The different contributions to the rotational energy for \ce{CS_2} look qualitatively similar to those for \ce{I_2}. As expected, $\langle 2 \mathbf{\Lambda \cdot L}\rangle$ deviates from $\langle \mathbf{L^2} \rangle$ at smaller $L$ for \ce{CS_2} due to its larger rotational constant.

\section{Assymptotic behaviour of $\langle \mathbf{\Lambda ^2} \rangle$}

\begin{figure}[h!]
\includegraphics[width=0.5\linewidth]{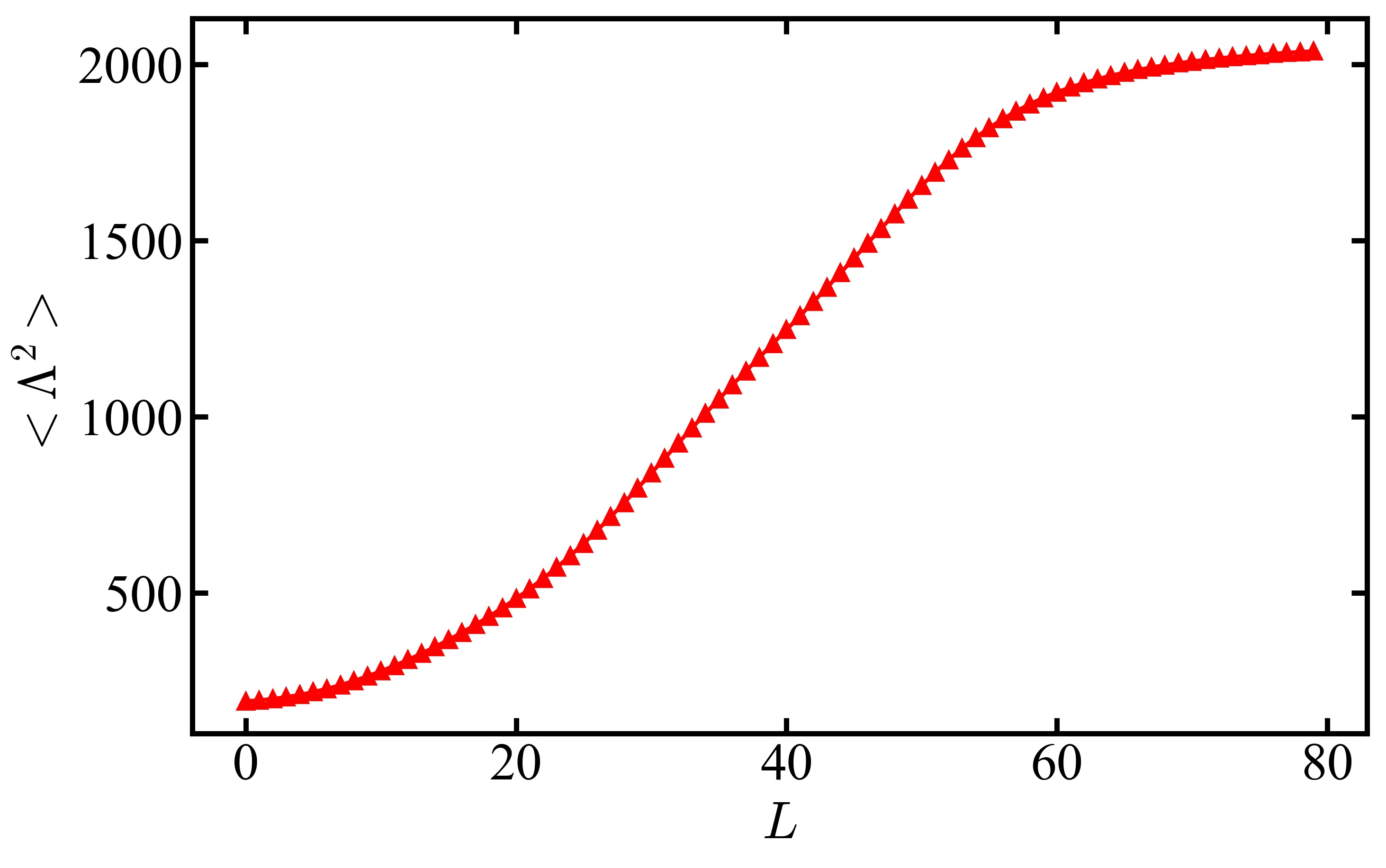}
\caption{$\langle \mathbf{\Lambda ^2} \rangle $ as a function of $L$ for \ce{I_2}}
\label{s3}
\end{figure}

\Autoref{s3} shows $\langle \mathbf{\Lambda ^2} \rangle $ for \ce{I_2}. In the limit $L \to \infty$ $\langle \mathbf{\Lambda ^2} \rangle $ converges to $\approx (3 \lambda)^2$ which is the maximum rotational energy a triple excitation can carry. 

\section{Parameters of the model}

The parameters used for simulations of molecular spectra in helium droplets are listed in Table \ref{Tab:Tcr}.
\par
$u$ characterises the anisotropy of the molecule-heilum interaction. To obtain an estimate of $u$, we use PES of a $\mathrm{molecule~+~^4He}$ complex and evaluate another anisotropy parameter $\Delta$ as defined in Ref. \cite{LemeshkoDroplets16_sup}

\begin{equation}
\Delta = \sqrt{\frac{5}{4\pi}}\frac{ \vert V^{\parallel}_{\mathrm{eff}}  - V^{\perp}_{\mathrm{eff}}\vert }{2}
\end{equation}
where $V^{\parallel}$ and $V^{\perp}_{\mathrm{eff}}$ are chosen as average potential well depth in the linear and $T$-shape geometries, respectively. One of the possible options is to fix $u$ such that the $B^*/B$ ratio is reproduced for one of the molecules and to scale it linearly with $\Delta$ for other molecules. Further fine-tuning of $u$ allows us to obtain a good agreement with experimental data.
\par
A similar strategy was applied to $\lambda$. In a semiclassical picture, this parameter plays a role of the number of He atoms in the solvation shell surrounding the molecule, thereby it reflects the angular symmetry of the interaction \cite{LehmannJCP01_sup}. $\lambda$ was fixed to 14 for $\ce{I_2}$. $\ce{CS_2}$ showing larger renormalization requires larger $\lambda$.
\par
Finally, we fix the frequency of the bosonic bath to the roton energy of bulk $^4$He $\omega = \SI{6}{cm^{-1}}$ \cite{Donnelly1981}. Roton-rotation coupling is known to give the dominant contribution to the molecule-helium interaction due to high density of states in the vicinity of the roton mode \cite{Zillich2004_sup}.

\begin{table}[h]
\setlength\extrarowheight{2pt}
\setlength{\tabcolsep}{5pt}
\begin{tabular}{|c|c|c|c|c|}
\hline
\multirow{2}{*}{Molecule} & \multirow{2}{*}{$B$ (\ce{GHz})} & \multirow{2}{*}{$u$ (\ce{GHz})} & \multirow{2}{*}{$\Delta$ (\ce{GHz})} & \multirow{2}{*}{$\lambda$} \\
             &      &      &      &    \\ \hline
$\ce{CS_2}$  & 3.27 \cite{Walker1971} & 1470 & 209 \cite{Prosmiti2009, Kalemos2012, Farrokhpour2013} & 24 \\ \hline
$\ce{I_2}$   & 1.12 \cite{Barrow1973} & 390 & 71 \cite{Garcia-Gutierrez2009, Limin2014} & 14 \\ \hline
\end{tabular}
\caption{Parameters used in calculations of rotational spectra.\label{Tab:Tcr}}
\end{table}

\end{document}